# Software-Defined Silicon Photonics based Metro Node for Spatial and Wavelength Superchannel Switching

Vidak Vujicic, Aravind P. Anthur, Alexander Gazman, Colm Browning, M. Deseada
Gutierrez Pascual, Ziyi Zhu, Keren Bergman and Liam P. Barry

**Abstract— Due to the growing popularity of optical
superchannels and software defined networking,
reconfigurable optical add-drop multiplexer
(ROADM) architectures for superchannel switching
have recently attracted significant attention.
ROADMs based on micro electro-mechanical system
(MEMS) and liquid crystal-on-silicon (LCoS)
technologies are predominantly used. Motivated by
requirements for low power, high-speed, small area
footprint and compact switching solutions, we
propose and demonstrate spatial and wavelength
flexible superchannel switching using monolithically
integrated silicon photonics (SiP) micro-ring
resonators (MRR). We demonstrate the MRRs
capabilities and potential to be used as a
fundamental building block in ROADMs. Unicast and
multicast switching operation of an entire
superchannel is demonstrated after transmission
over 50 km of standard single mode fiber. The
performance of each sub-channel from the 120 Gb/s
QPSK Nyquist superchannel is analyzed and
degradation in error vector magnitude performance
was observed for outer sub-channels due to the 3-dB
bandwidth of the MRRs, which is comparable with
the superchannel bandwidth. However, all sub-
channels for all switching cases (unicast, multicast
and bi-directional operation) exhibit performance far
below the 7% FEC limit. The switching time of the SiP
MRR chip is such that high capacity superchannel
interconnects between users can be setup and
reconfigured on the microsecond timescale.**

resonators, Reconfigurable Add-Drop Multiplexer
(ROADM), Silicon Photonics, Superchannel,
Wavelength Selected Switch (WSS).

V. Vujicic, A. P. Anthur, C. Browning, M. D. G. Pascual and L.
P. Barry are with the Department of Electronic Engineering,
Dublin City University, Glasnevin, Dublin 9, Ireland
(vidak.vujicic@dcu.ie).
A. Gazman, Z. Zhu and K. Bergman are with the Department of
Electrical Engineering, Columbia University, New York, New York
10027, USA.
M. D. G. Pascual is also with Pilot Photonics Ltd, Dublin 9,
Ireland.

## I. Introduction

Due to the continuing growth of residential, mobile,
business and cloud traffic, metro networks are
experiencing a significant transformation, resulting in
metro traffic growing about two times faster than traffic
going into the backbone by 2017 [1]. Moreover, a significant
percentage of the metro network traffic will be terminated
within the metro network itself, as traffic flows between
data-centers and end-user's premises for the delivery of
cloud services go via metro networks instead of traversing
through the backbone network [1]. Therefore, traffic
exchange between metro networks on one side, and data
centers and access networks on the other, will experience
massive growth over the coming decade. The estimated
traffic growth in conjunction with the requirements for
flexible traffic management will require scaling of metro
network nodes [2-4]. Scaling must occur on multiple levels
in order to support increased total capacity of metro
networks, while enabling a higher degree of flexibility and
reconfigurability with fast reconfiguration times [2-5].

Optical superchannels are a proven technique for
increasing the total capacity of optical systems by enabling
efficient utilization of the available spectrum [6-9].
Proposed superchannel architectures in core, metro and
access networks [6,10,11] have motivated research on the
implementation of superchannel spatial switching using
wavelength selective switches (WSS), which are the main
building blocks of reconfigurable add-drop multiplexer
(ROADM) nodes [12-15]. Whilst various underlying
technologies for WSSs have been proposed, such as
interferometric spatial switching [16] and spatial beam
splitting [17], switching architectures based on micro
electro-mechanical system (MEMS) and liquid crystal-on-
silicon (LCoS) are the most mature, and currently available
commercial WSS devices are based on those technologies
[18-21]. Most WSSs are based on free-space optics, such as
the case with MEMS and LCoS based WSSs, mainly due to
flexibility in filtering bandwidth and the high degree of
parallelism of free-space optics [22]. Furthermore, flexible
optical signal processing can be achieved in free-space
optics by using spatial light modulators [22]. Nevertheless,
the three-dimensional structure of free-space optics is not
suitable for realizing small size devices via integration.
Also, the switching times of free space optics based WSSs is in the order of milliseconds, which makes them efficient only for large and steady traffic flows [23].

Integrated silicon photonics (SiP) based WSSs are an attractive alternative for realizing high-speed optical switches with small area footprint, low power consumption and the potential for reduced fabrication costs at large scales [23-26]. Recently, a two-dimensional silicon integrated MEMS device with switching times in the order of microseconds has been proposed [27]. However, a drawback of MEMS-based switches is that they typically require large driving voltages [23]. Two other popular structures for optical switching elements are micro-ring resonators (MRR) [23] and Mach-Zehnder (MZ) interferometers [24]. Although MZ based switches do not require wavelength tuning like MRR switches [24], the advantage of MRR over MZ switches, in terms of power consumption, has previously been demonstrated [23]. The WDM switching capabilities have been demonstrated for MRR based switch fabrics for several modulation formats intended for data-center applications [28,29]. However, due to the increasing popularity of superchannel architectures, it is necessary to investigate switching performance of MRR based WSSs in a superchannel based optical system.

In this paper, we demonstrate for the first time to the best of our knowledge, superchannel spatial and wavelength switching using monolithic SiP MRRs. A programmable SiP chip, based on eight cascaded MRRs is employed to perform reconfigurable wavelength routing (unicast switching), as well as optical multicasting. The MRR switch is operational over the entire C-band, which is demonstrated by investigating the superchannel switching performance at three different center wavelengths across the band. The 120 Gb/s superchannel consists of six 20 Gb/s QPSK Nyquist filtered sub-channels. The switching functionality is demonstrated after transmission over 50 km of standard single-mode fiber (SSMF). We also demonstrate bi-directional operation of the MRR switch. The rest of this paper is organized in four sections. Section 2 presents the proposed SiP MRR chip and envisioned metro ROADM node architecture. Section 3 steps through the experimental setup, whilst Section 4 presents experimental results. Finally, conclusions are drawn in Section 5.

## II. PROGRAMMABLE SILICON PHOTONIC CHIP

The SiP chip used for this work consists of eight cascaded MRRs, with one input port and eight output ports as depicted in Fig. 1. This configuration acts as a reconfigurable 1×8 spatial switch, as each MRR exhibits wavelength selectivity by virtue of the fact that their passbands may be thermally tuned in order to drop an incoming wavelength at a desired point. Additionally, the MRR switch may operate in a multicast regime when passbands are tuned to overlap so that the same incoming wavelength is dropped at multiple ports. This operation is provided by the spectral Lorentzian response of each MRR [30], whose alignment enables ideally equal power distribution of the optical signal among the desired output ports. However, due to the imperfections in optical fiber coupling into the SiP chip, the power of the optical signal among the desired output ports may vary by a few dB. The switching time of the SiP MRR chip is such that unicast and multicast functionality can be setup and reconfigured on the microsecond timescale [23].

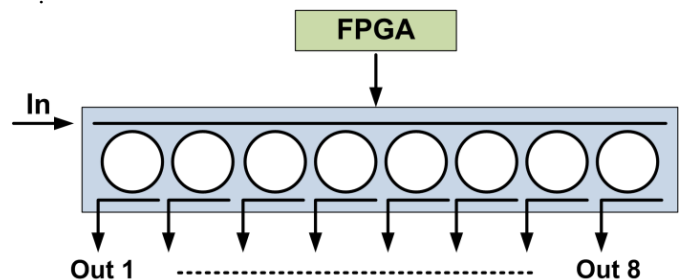

Fig. 1. Block diagram of the SiP MRR device. FPGA is used to reconfigure the SiP device to switch the data to any of the eight ports for unicast and multicast operation.

The MRR switch is operational over the entire C-band. The operational principle of the MRRs is based on the thermo-optic effect [31]. The measured thermo-optic response of each individual ring is stored in the control plane and utilized for the effective tuning of the rings' passbands. Fig. 2(a) shows normalized wavelength shift as a function of increased bias voltage. The rings are designed such that their center frequencies are separated by 1.27 nm (~ 160 GHz) with an FSR of 13 nm, as shown in Fig. 2(b). An increase in the supplied voltage to the heater causes a change in the effective index of the optical mode via an increase in the local temperature of the ring, allowing the resonance of the rings to shift to higher wavelengths. A continuous and accurate relationship between the supplied bias voltage and the frequency response of the MRR chip is enabled by polynomial fitting (2nd order) of the measured results. The fitted results are used together with the MRRs' Lorentzian resonant response and zero bias resonances to develop dynamic models to perform any type of switching operation [32].

The detailed description of the control plane and software implementation is given in [32]. The switching algorithm used to configure the SiP chip for all possible switching cases is described in details in [32]. The switching algorithm is implemented in Python programming language and post calculation commands are sent to an Altera FPGA that controls the SiP MRR device. The algorithm takes two arguments. The first is wavelength of the input channel, and the second argument is a binary array indicating where to route the signal. For instance, a binary array given as [0, 0, 1, 0, 0, 0, 0, 0] would route the input signal to output port 3, whilst an array [0, 0, 0, 0, 1, 1, 1, 0] would initiate multicast operation on output ports 5, 6 and 7. The software calculates in a power efficient manner the required wavelength shifts for the MRR array to perform the required switching task [32]. The calculated wavelength shifts are converted to the corresponding bias voltages and sent to the control FPGA of the SiP device. The FPGA board is connected to digital-to-analog converters (DAC) to provide voltages required for thermal tuning of the MMRs in order to achieve the desired functionality − unicast or multicast. The pre-determined signals from the DACs are amplified and buffered to allow a tuning range over the C-band. Due to the symmetrical spectra of the MRRs it is possible to drop the required amount of optical power from either side of the Lorentzian response. To optimize the tuning power consumption of the MRR array the software automatically selects the closest tuning option [32]. The control plane was designed to automatically enable wavelength selection using only desired passband. The automatic selection of a desired passband is enabled by calculating required bias voltages for the specified wavelength (resonance) and by detuning undesired resonances to ensure maximized power drop for unicast operation and uniform power distribution among channels for multicast transmission [32].

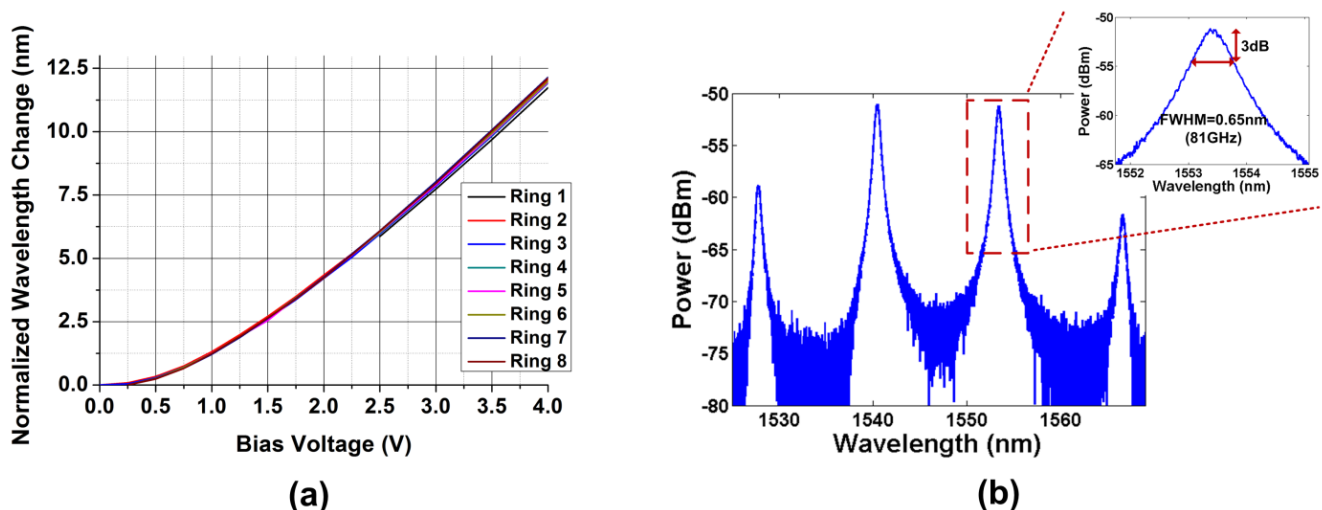

Fig. 2. (a) Measured thermo-optic response of the heater-ring system of the eight rings. (b) Measured response profile of the MRR at drop port 1 (Resolution bandwidth 1.44 pm). The inset shows the measured 3-dB bandwidth of drop port 1 at 1553.5 nm.

The MRRs are designed each with a varying diameter around 14 μm, enabling different wavelengths of operation when no voltage is applied [30, 32]. As shown in Fig. 2(b), the 3-dB bandwidth of MRRs is estimated to be in the range 76-86 GHz, for wavelengths covering the C-band. Several parameters in the physical design of the MRRs can affect the 3-dB bandwidth variation of each ring, including ring circumference imperfections and wavelength dependence of the effective refractive index [33]. However, the coupling coefficient is the dominant factor causing 3-dB bandwidth variation in a range of a few GHz [33], which can be justified by the gap of around 100 nm between the ring and the main waveguide. The MRR chip has been designed with large gaps between rings in order to prevent thermal crosstalk between rings. It is experimentally verified that increased bias on one or several rings does not affect the spectral response of the neighboring rings. The MRR chip does not have a hitless function [34]. Due to fiber grating alignment issues, the chip exhibits ~20 dB of loss. However, it was shown that in the case of effective fiber coupling, it would be possible to achieve significantly lower losses in the SiP MRR chip [35]. The tuning efficiency of the MRR device is 0.266 nm/mW [32]. Therefore, the energy consumption required for reconfiguration of the silicon photonic device for all possible wavelength operations and output ports combinations for unicast and multicast functions is estimated to be from 1.25 to 96.8 fJ/bit, depending on switching operation [32].

## A. Envisioned Metro ROADM Node Architecture

In light of increased cloud traffic between data centers and end users, it has been proposed that future metro ROADM nodes should provide multicast functionality along with unicast traffic switching to resolve wavelength contention [2,36]. The majority of the currently implemented ROADM nodes suffer from lack of the flexibility in reconfiguring wavelength paths based on user's requirements. Colorless and directionless (CD) and colorless, directionless and contentionless (CDC) ROADMs allow flexible wavelength reassignment at the expense of higher node complexity and cost [37,38].

WSS modules within the current CD and CDC ROADMs are based on MEMS and LCoS technology, providing bandwidth flexibility, generation of arbitrary passband profiles and high number of add-drop ports. However, the bulky structure, high power consumption and millisecond switching times of standard WSSs could be improved by the utilization of SiP MRR switching fabric. Figure 3 illustrates the envisioned structure and functionality of a metro ROADM node where 1xN SiP switching fabric is used as 1xN WSS, capable of performing unicast and multicast switching operation. The illustrated metro ROADM node provides colorless, directionless and contentionless functionality [37-39], which is enabled by the MRR's inherent characteristics such as wavelength flexibility and bi-directionality. Bi-directionality provides the opportunity of using the same switching fabric for all traffic directions, which potentially reduces fabrication costs at large scales. Compared to the standard CDC ROADMs where multicast functionality is provided only at multicast MxN add-drop switches [37,38], in the case of SiP based ROADM multicast functionality is available at each 1xN switch and at MxN add-drop switch aswell. Moreover, as some vendors consider superchannels as a single unit of capacity seen by the services that use it, spatial switching of an entire superchannel would be required [40,41]. The switching matrix elements would be based on MRR switching fabric as well, as illustrated in Fig. 3 [23,39,42,43]. Even though MRR based WSS cannot provide arbitrary passbands such as LCoS based switches and select a number of independent wavelengths on the same port, the switching matrix would route the desired independent wavelengths from different output ports to a selected WSS input port [39]. It should be noted that for the practical implementation of an MRR based ROADM, further performance improvement is required in terms of loss reduction [23,32], polarization insensitivity [39], bandwidth flexibility [33], thermal stabilization [44] and the design of the MxN MRR switch [39]. Therefore, further research is required for the MRRs to reach the specifications achieved by the current WSS. A full flexibility in customizing the number of sub-channels and total superchannel bandwidth would be possible with bandwidth-tunable MRRs [33]. Furthermore, for certain applications where superchannels consist of a high number of high-bandwidth sub-channels, superchannel sub-band filtering would be required as well, which necessitates MRR switches with a sharper roll-off factor - potentially facilitated by higher-order MRR switches [12,39].

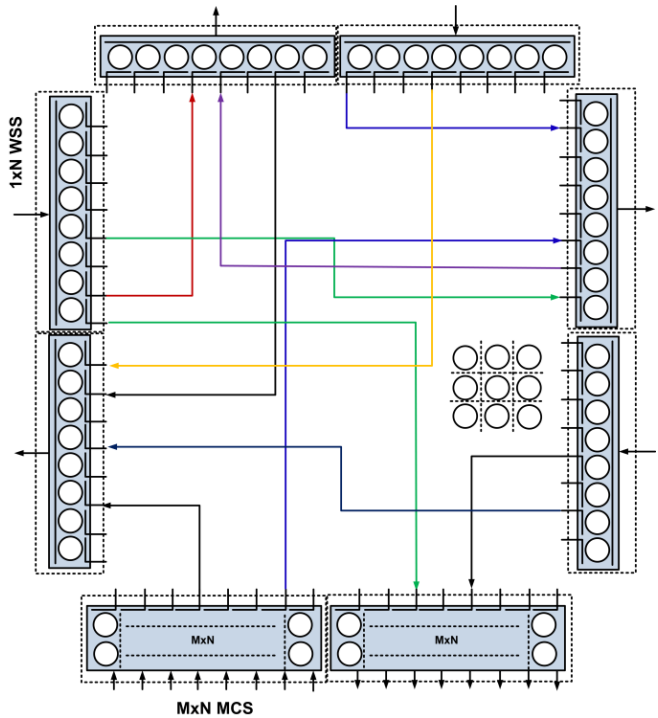

Fig. 3. Envisioned metro network CDC ROADM node architecture with unicast and multicast operation based on MRR switches. MCS – Multicast Switch.

## III. EXPERIMENTAL CONFIGURATION

The experimental setup of the proposed system for superchannel switching using MRRs is shown in Fig. 4. At the transmitter side, we use a tunable gain-switched optical comb source for the wavelength-flexible superchannel generation. The detailed configuration of the tunable comb source, which is based on an injection-locked gain-switched Fabry-Perot laser diode (FP-LD) is presented in [45]. A tunable laser with an optical output power of 9-12 dBm and <100 kHz estimated linewidth acts as the master and injects light into the slave FP-LD via a circulator. The wavelength of the master is tuned to match one of the longitudinal modes of the FP-LD, which enables injection locking on a particular longitudinal mode of the FP laser, so the other modes are suppressed and single-mode operation of the FP laser is achieved at the selected wavelength. The FP-LD is then gain-switched with the aid of a 12.5 GHz RF sinusoidal signal for the generation of an optical comb. The optical comb source was set at three different wavelengths of 1539 nm, 1552 nm and 1563 nm in order to test the superchannel switching system. At each operating wavelength, six comb lines within a 3-5 dB flatness window, and frequency spaced by 12.5 GHz are generated. These comb lines are selected with the aid of a waveshaper (WS), prior to being amplified with an Erbium doped fiber amplifier (EDFA). The filtered comb tones are modulated by a single complex Mach-Zehnder modulator (IQ-MZM). The IQ-MZM is modulated with an amplified 10 GBaud Nyquist

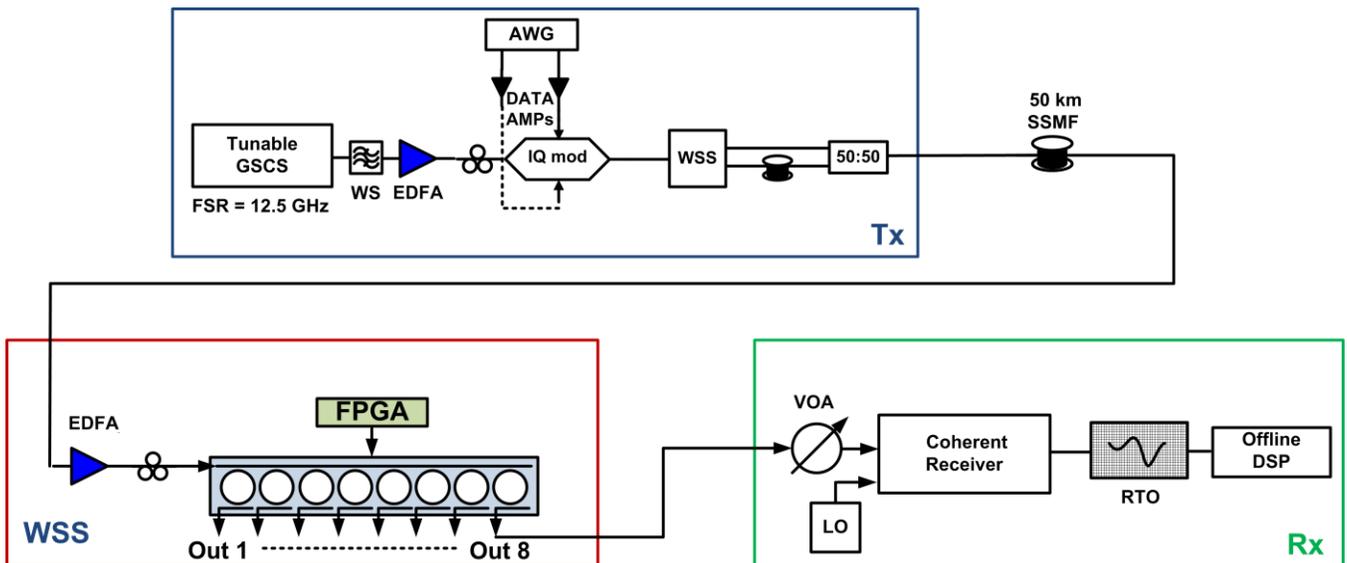

Fig. 4. Schematic of the experimental setup. Gain-Switched Comb Source (GSCS); Waveshaper (WS); Polarization Controller (PC); Arbitrary Waveform Generator (AWG); IQ Mach Zehnder Modulator (IQ-MZM); Data Amplifier (Data Amp); Erbium Doped Fiber Amplifier (EDFA); Variable Optical Attenuator (VOA); Local Oscillator (LO); Real Time Oscilloscope (RTO), Field Programmable Gate Array (FPGA).

QPSK signal waveform derived from an arbitrary waveform generator (AWG, Keysight M8195A) operating at 60 GSa/s. A digital root raised cosine filter with a roll-off factor of 0.1% is applied to the QPSK signal. Therefore, the bandwidth of the optical Nyquist QPSK signal is ~10 GHz. The total bandwidth of the superchannel is ~85 GHz. The sub-channels of the superchannel signal are de-correlated using an odd and even configuration with the WSS and 10 m SSMF patchcord in one branch. The de-correlated signal is then transmitted over 50 km of SSMF.

After transmission over 50 km SSMF the signal is first amplified with an EDFA and then directed to the SiP MRR chip. The amplification stage in this case is used because of the loss exhibited by the chip due to the fiber grating alignment issues. An FPGA is used to control the switching operation of the MRRs, and tune one (unicast) or more (multicast) rings to drop the incoming superchannel. At the receiver side, the phase diversity coherent receiver is used to perform coherent detection. The optical local oscillator (LO) is provided by an ECL that has 13 dBm output power and <80 kHz optical linewidth. The LO is tuned to each of the 6 sub-channels from a superchannel for selection and detection. The LO at the coherent receiver negates the need for an optical filter to select out desired channel. The received signal is captured with a real time oscilloscope (RTO, Tektronix MSO71254C) operating at 50 GSa/s. Digital processing of the received signal (timing deskew, resampling, adaptive equalization, matched filtering, frequency offset compensation, carrier phase recovery and phase tracking), and error vector magnitude (EVM) calculations, are performed offline using Matlab.

## IV. RESULTS AND DISCUSSION

The obtained experimental results for MRR unicast operation are shown in Fig. 5, where the superchannel is switched to one of the eight output ports of the MRR and the EVM is estimated for each sub-channel. Figures 5(a)-(c) show the results for the superchannel switching at 1539 nm wavelength, whilst Fig. 5(d)-(f) and Fig. 5(g)-(i) show results for the superchannel switching at 1552 nm and 1563 nm wavelength respectively.

The optical spectra of the externally injected GSCS at 1539 nm, 1552 nm and 1563 nm are shown in Fig. 5(a), 5(d) and 5(g) respectively. The six comb tones within a flatness of 3-5 dB are selected and used for superchannel generation, as illustrated in Fig. 5(b), 5(e) and 5(h). The corresponding optical carrier-to-noise ratio (OCNR), measured prior to amplification at the transmitter (see Fig. 4), for all comb lines used for the data transmission experiments is found to be greater than 49 dB (RBW=1.44 pm) at 1539 nm, 52 dB at 1552 nm and 50 dB at 1563 nm.

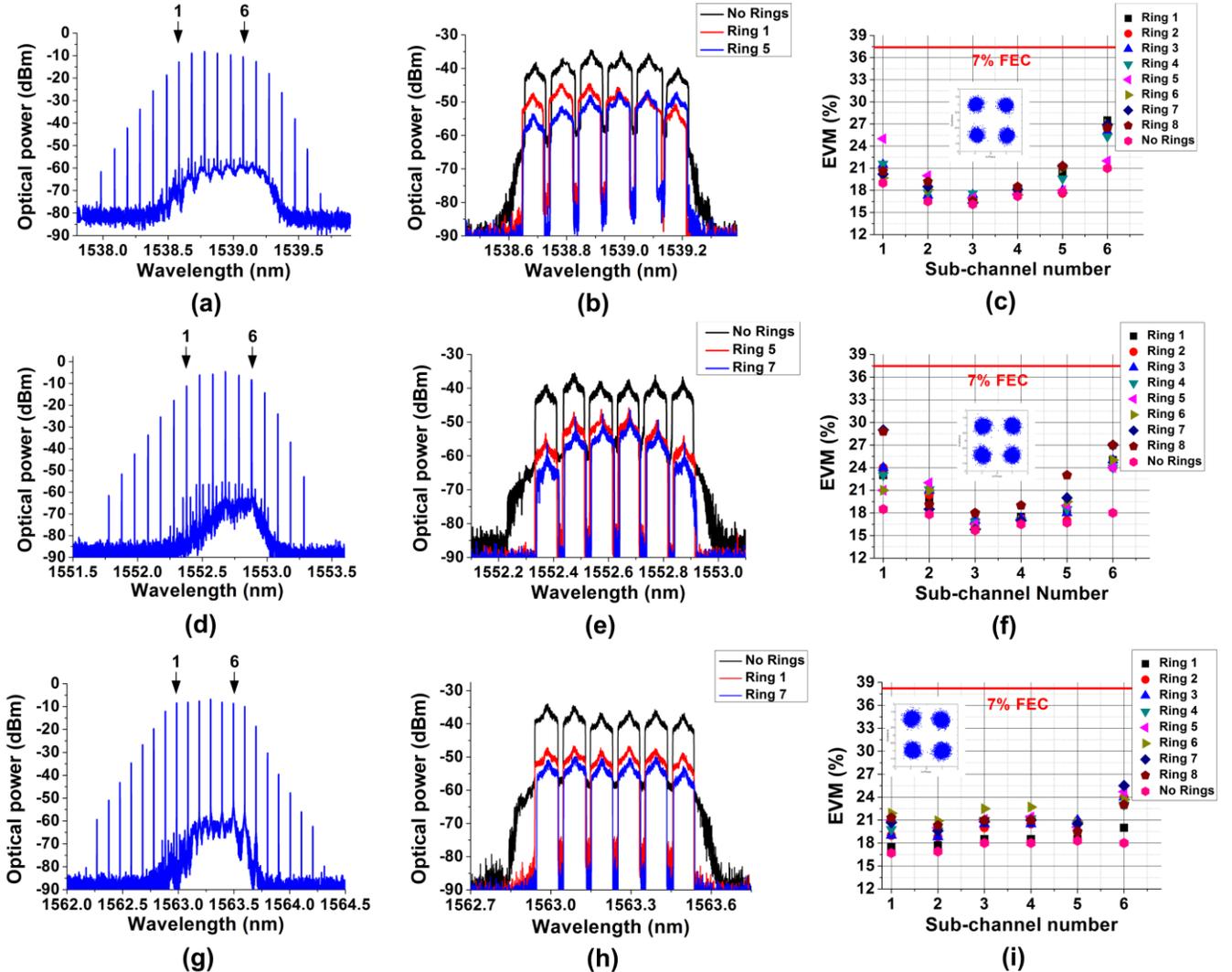

Fig. 5. a) Optical spectrum of externally injected GSCS with 12.5 GHz FSR (RBW=1.44 pm) at 1539 nm central wavelength; (b) Optical spectrum of 1539 nm superchannel before and after MRR switch (RBW=1.44 pm); (c) Measured EVM for each ring and each sub-channel for unicast switching, after transmission over 50 km of SSMF. Inset shows a measured constellation diagram. Fig. 5. (d-i) show the same results as (a-c), but for the superchannels at 1552 nm and 1563 nm, respectively.

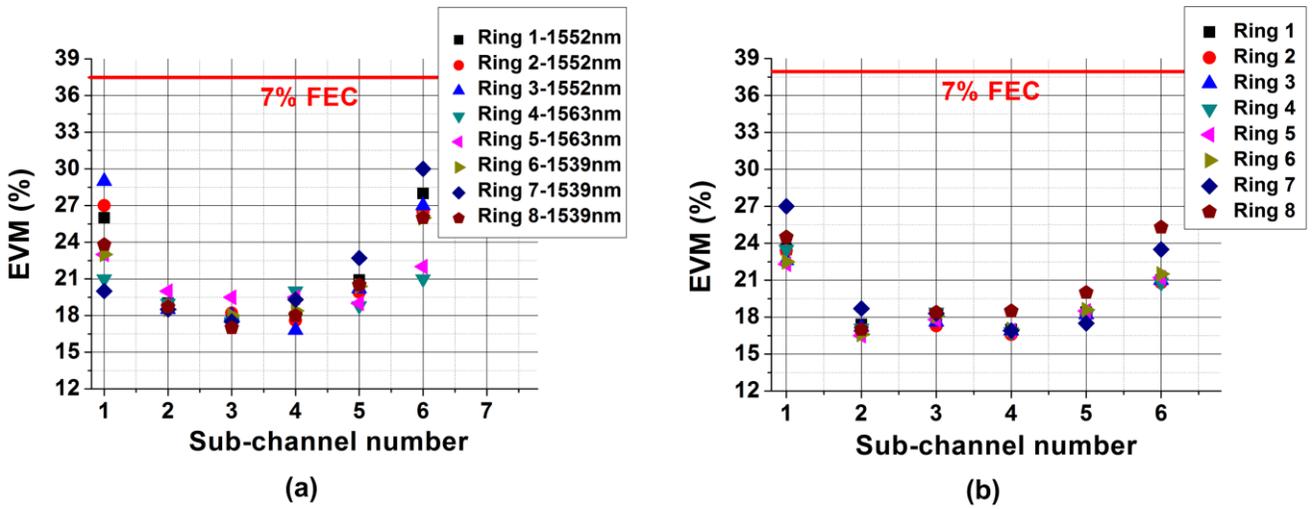

Fig. 6. (a) Measured EVM performance for multicast switching for each ring and sub-channel operating at three different wavelengths. (b) Measured EVM performance for bi-directional unicast switching at 1552 nm.

Figure 5(b) illustrates the superchannel optical spectra at 1539 nm before and after the MRR switch. Channel performance is determined using EVM measurements for a total received optical power (received power of the entire superchannel) of ~ -25 dBm, and results obtained for the 1539 nm superchannel are given in Fig. 5(c). After transmission over 50 km of SSMF and passing through the MRR, all channels exhibit performance far below the 7% FEC limit (EVM=38% corresponding to BER=3×10⁻³ [46]). Due to the MRRs 3–dB bandwidth in the range of 80-85 GHz at the 1539 nm wavelength and superchannel bandwidth of 85 GHz, performance degradation of the outer channels occurs as shown in Fig. 5(c). The effect of attenuation of the outer channels due to the bandwidth constrains is visible in Fig. 5 (b), where the passband of Ring 1 of the MRR device is slightly detuned towards lower wavelengths from the central wavelength. This causes sub-channel 1 (black squares in Fig. 5(c)) to exhibit slight degradation compared to the case when the MRR is not used, whilst sub-channel 6 has higher degradation compared to the case when the MRR is not used. However, if the detuning is towards higher wavelengths, as shown in Fig. 5(b); such as is the case for Ring 5, sub-channel 1 (pink triangles in Fig. 5(c)) will exhibit a penalty in performance while sub-channel 6 will perform similarly to the case when the MRR is not used. The measured constellation diagram for sub-channel 3 at Ring 1 is shown as an inset in Fig. 5(c).

All sub-channels within 1552 nm superchannel exhibit performance below the 7% FEC limit, as shown in Fig. 5(f). The performance is measured after transmission over 50 km of SSMF, and then passing through the SiP MRR which is tuned to the correct wavelength. However, due to the slightly lower bandwidth of some rings at the 1552 nm wavelength, performance degradation of outer sub-channels occurs. The measured 3-dB bandwidth at 1552 nm wavelength for all MRRs is in range 76-85 GHz, so sub-channels 1 and 2, as well as 5 and 6 exhibit performance degradation which is higher than in the case of the 1539 nm superchannel. This effect is visible in Fig. 5 (e) and (f). The measured constellation diagram for sub-channel 5 at Ring 4 is shown as an inset in Fig. 5(f).

In accordance with the previous results, after transmission over 50 km of SSMF and passing through the MRR, satisfactory performance was achieved on all sub-channels at 1563 nm. Due to the MRRs 3–dB bandwidth in the range ~82-86 GHz at the 1563 nm wavelength, performance degradation of outer sub-channels is lower compared to the previous case. The measured constellation diagram for sub-channel 5 at Ring 1 is shown as an inset in Fig. 5(i).

The results obtained for multicast operation are given in Fig. 6(a). Multicast functionality with the SiP chip is implemented by simultaneous thermal tuning of the passbands of up to three rings to overlap such that, ideally, equal power was dropped from each port. However, due to the non-ideal response of the MRR to the applied voltages, the dropped powers at different ports may vary by 1-2 dB. As the optical signal is now split into three paths rather than one, the output power from each ring drops by at least 4.8 dB. Nevertheless, the received power was sufficient to perform measurements at -25 dBm, similar to the unicast case. Multicast operation is demonstrated for rings 1 to 3 at 1539 nm, rings 4 and 5 at 1552 nm and rings 6 to 8 at 1563 nm. Although sub-channels 1 and 6 exhibit higher degradation compared to unicast switching case, all sub-channels demonstrate satisfactory performance - below the 7% FEC limit. The performance difference compared to the unicast switching is caused by a slight 3-dB bandwidth reduction of MRR passbands when multiple rings are tuned simultaneously, due to the coupling effects between the main waveguide and drop rings.

The potential for bi-directional operation of the MRR is confirmed by passing the superchannel signal after transmission through 50 km of fiber to one of the 8 output ports of the MRR switch (which serves as an input port in this case). The superchannel signal is detected at the 'input' port (which serves as an output port in this case). The performance of the bi-directional unicast operation at 1552 nm is shown in Fig. 6(b). As similar performance to the unicast switching (see Fig. 5(f)) of the superchannel at different rings was obtained, the possibility of bi-directional operation of the MRR chip is demonstrated.

## V. Conclusion

Due to the rapidly growing traffic in metro networks caused by the increase in cloud services, metro rings are experiencing a significant transformation which will require significant scaling of metro network nodes. Future metro ROADM nodes should support increased total capacity of metro networks, enable a higher degree of flexibility and reconfigurability and provide faster reconfiguration times. In this paper we investigate, for the first time, the possibility of utilizing monolithically integrated silicon photonic micro-ring resonators for spatial and wavelength switching of entire superchannels. A programmable SiP chip, based on eight cascaded MRRs, is employed to perform reconfigurable wavelength routing (unicast switching), as well as optical multicasting. The MRR switch is operational over the entire C-band, which is demonstrated by investigating the entire superchannel switching performance at three different wavelengths. The 120 Gb/s superchannel consists of six 20 Gb/s QPSK Nyquist filtered sub-channels. The switching functionality is demonstrated and analyzed after transmission over 50 km of SSMF. As the 3-dB bandwidth of the MRRs is in the range 76-86 GHz, and the superchannel bandwidth is 85 GHz, outer sub-channels within the superchannel exhibit performance degradation compared to the middle sub-channels of the superchannel signal. However, all sub-channels exhibit satisfactory EVM performance which is measured to be far below 7% FEC limit after transmission over 50 km of SSMF and switching via the MRRs. To avoid degradation of outer sub-channels, it is possible to design MRRs with larger 3-dB bandwidth. We also demonstrate bi-directional operation of the MRR switch which demonstrates the opportunity of using the same switching fabric for all traffic directions, which potentially could reduce fabrication costs of Si MRR based ROADMs. Advances in silicon photonics will lead to improvements in terms of loss reduction, bandwidth flexibility and integration of an MxN MRR matrix, adding to its suitability for inclusion in future metro ROADM nodes by providing a high degree of flexibility and programmability. Furthermore, the design of higher order MRRs would increase networking granularity by enabling sharper passband profiles – facilitating MRR's intra-superchannel filtering.

## Acknowledgment

This work was jointly supported through the US-Ireland (15/US-C2C/I3132), EU FP7 "BIGPIPES", CONNECT (13/RC/2077), IPIC (12/RC/2276), CIAN NSF ERC (EEC-0812072) and NSF (CNS-1423105) research grants. This material is based on research sponsored by Air Force Research Laboratory under agreement number FA8650-15-2-5220. The U.S. Government is authorized to reproduce and distribute reprints for Governmental purposes notwithstanding any copyright notation thereon. This work was supported by grant DE- SC0015867 through Advanced Scientific Computing Research within the Department of Energy Office of Science. The authors would also like to acknowledge the support of Freedom Photonics (subcontract CU 16-0418) and Pilot Photonics Ltd for use of their Tunable Comb Source.

## References

[1] Bell Labs, "Metro Network Traffic Growth: An Architecture Impact Study", Strategic White Paper, Dec. 2013.

[2] M. Schiano, A. Percelsi, and M. Quagliotti, "Flexible Node Architectures for Metro Networks [Invited]," IEEE/OSA J. Opt. Commun. Networking, vol. 7, no. 12, pp. B131-B140, Dec. 2015.

[3] S. Gringeri, B. Basch, V. Shukla, R. Egorov, and T. J. Xia, "Flexible Architectures for Optical Transport Nodes and Networks," IEEE Commun. Mag., pp. 40-50, July 2010.

[4] M. Cvijetic, I. B. Djordjevic, and N. Cvijetic, "Multidimensional Elastic Routing for Next Generation Optical Networks," 13th Intern. Conf. on High Performance Switching and Routing, DOI: 10.1109/HPSR.2012.6260850, Jun 2012.

[5] M. Nooruzzaman and E. Halima, "Low-Cost Hybrid ROADM Architectures for Scalable C/DWDM Metro Networks," IEEE Commun. Mag., pp. 153-161, Aug. 2016.

[6] J. Pfeifle, V. Vujicic, R. T. Watts, P. C. Schindler, C. Weimann, R. Zhou, W. Freude, L. P. Barry, C. Koos, "Flexible terabit/s Nyquist-WDM super-channels using a gain-switched comb source," Opt. Express, vol. 23, no. 2, pp. 724-738, Jan. 2015.

[7] P. Marin-Palomo, J. N. Kemal, M. Karpov, A. Kordts, J. Pfeifle, M. H. P. Pfeiffer, P. Trocha, S. Wolf, V. Bracsch, R. Rosenberger, K. Vijayan, W. Freude, T. J. Kippenberg, C. Koos, "Microresonator solitons for massively parallel coherent optical communications," arXiv:1610.01484 [nlin.PS], https://arxiv.org/pdf/1610.01484v2.pdf

[8] V. Ataie, E. Temprana, L. Liu, E. Myslivets, B. P. P. Kuo, N. Alic, and S. Radic, "Ultrahigh Count Coherent WDM Channels Transmission Using Optical Parametric Comb-Based Frequency Synthesizer," J. Lightwave Technol., vol. 33, no. 3, Feb. 2015.

[9] C. Lin, I. B. Djordjevic, M. Cvijetic, and D. Zou, "Mode-Multiplexed Multi-Tb/s Superchannel Transmission With Advanced Multidimensional Signaling in the Presence of Fiber Nonlinearities," IEEE Trans. Commun., vol. 62, no. 7, Jul. 2014.

[10] T. Shao, R. Zhou, V. Vujicic, M. D. Gutierrez Pascual, P. M. Anandarajah, and L. P. Barry, "100 km Coherent Nyquist Ultradense Wavelength Division Multiplexed Passive Optical Network Using a Tunable Gain-Switched Comb Source," IEEE/OSA J. Opt. Commun. Networking, vol. 8, no. 2, pp. 112-117, Feb. 2016.

[11] R. Maher, A. Alvarado, D. Lavery and P. Bayvel, "Increasing the information rates of optical communications via coded modulation: a study of transceiver performance," Scientific Reports, vol. 6, article number: 21278, Feb. 2016.

[12] L. Zhuang, C. Zhu, B. Corcoran, M. Burla, C. G. H. Roeloffzen, A. Leinse, J. Schröder, and A. J. Lowery, "Sub-GHz-resolution C-band Nyquist-filtering interleaver on a high-index-contrast photonic integrated circuit," Opt. Express, vol. 24, no. 6, pp. 5715-5727, Mar. 2016.

[13] J. Li, M. Karlsson, P. A. Andrekson, and K. Xu, "Transmission of 1.936 Tb/s (11 × 176 Gb/s) DP-16QAM superchannel signals over 640 km SSMF with EDFA only and 300 GHz WSS

channel," Opt. Express, vol. 20, no. 26, pp. B223-B231, Nov. 2012.

[14] J. Renaudier, R. Rios Müller, L. Schmalen, P. Tran, P. Brindel and G. Charlet, "1-Tb/s PDM-32QAM Superchannel Transmission at 6.7-b/s/Hz over SSMF and 150-GHz-Grid ROADMs," in Proc. ECOC, Cannes, paper Tu.3.3.4, Sep. 2014.

[15] S. Kumar, R. Egorov, K. Croussore, M. Allen, M. Mitchell, B. Basch, "Experimental Study of Intra- vs. Inter-Superchannel Spectral Equalization in Flexible Grid Systems," in Proc. OFC/NFOEC, Los Angeles, CA, paper JW2A.05, Mar. 2013.

[16] S. J. Fabbri, S. Sygletos, A. Perentos, E. Pincemin, K. Sugden, and A. D. Ellis, "Experimental Implementation of an All-Optical Interferometric Drop, Add, and Extract Multiplexer for Superchannels," J. Lightwave Technol., vol. 33, no. 7, pp. 1351-1357, Apr. 2015.

[17] K. Yamaguchi, Y. Ikuma, M. Nakajima, K. Suzuki, M. Itoh, and T. Hashimoto, "M x N Wavelength Selective Switches Using Beam Splitting By Space Light Modulators," IEEE Photonics J., vol. 8, no. 1, pp. 0600809, Feb. 2016.

[18] U. Arad, Y. Corem, B. Frenkel, V. Deich, J. Dunayevsky, R. Harel, P. Janosik, G. Cohen, and D. M. Marom, "MEMS Wavelength-Selective Switch Incorporating Liquid Crystal Shutters for Attenuation and Hitless Operation," in Proc. International Conference on Optical MEMS and Nanophotonics, Kanazawa, Japan, Aug. 2013.

[19] Y. Sakurai, M. Kawasugi, Y. Hotta, M. D. S. Khan, H. Oguri, K. Takeuchi, S. Michihata, and N. Uehara, "LCOS-Based 4x4 Wavelength Cross-Connect Switch For Flexible Channel Management in ROADMs," in Proc. OFC/NFOEC, Los Angeles, CA, paper OTuM4, Mar. 2011.

[20] L. E. Nelson, M. D. Feuer, K. Abedin, X. Zhou, T. F. Taunay, J. M. Fini, B. Zhu, R. Isaac, R. Harel, G. Cohen, and D. M. Marom, "Spatial Superchannel Routing in a Two-Span ROADM System for Space Division Multiplexing," J. Lightwave Technol., vol. 32, no. 4, pp. 783-789, Feb. 2014.

[21] S. Frisken, I. Clarke and S. Poole, "Technology and Applications of Liquid Cristal on Silicon (LCoS) in Telecommunications," in Optical Fiber Telecommunications Volume VIA: Components and Subsystems, Chapter 18, Academic Press, May 2013.

[22] K. Suzuki, K. Seno, and Y. Ikuma, "Application of waveguide/free-space optics hybrid to ROADM device," J. Lightwave Technol., vol. PP, no. 99, Aug. 2016.

[23] D. Nikolova, S. Rumley, D. Calhoun, Q. Li, Robert. Hendry, P. Samadi, and K. Bergman, "Scaling silicon photonic switch fabrics for data center interconnection networks," Opt. Express, vol. 23, no. 2, pp. 1159-1175, Jan. 2015.

[24] S. Nakamura, S. Yanagimachi, H. Takeshita, A. Tajima, T. Hino, and K. Fukuchi, "Optical Switches Based on Silicon Photonics for ROADM Application," IEEE Sel. Top. Quantum Electron., vol. 22, no. 6, pp. 3600609, Nov/Dec. 2016.

[25] Q. Li, D. Nikolova, D. M. Calhoun, Y. Liu, R. Ding, T. Baehr-Jones, M. Hochberg, and K. Bergman, "Single Microring-Based 2 × 2 Silicon Photonic Crossbar Switches," IEEE Photonics Technol. Lett., vol. 27, no. 18, pp. 1981-1984, Sep. 2015.

[26] H. Kawashima, K. Suzuki, K. Tanizawa, S. Suda, G. Cong, H. Matsuura, S. Namiki, and K. Ikeda, "Multi-port Optical Switch Based on Silicon Photonics," in Proc. OFC/NFOEC, Anaheim, CA, paper W1E.6, Mar. 2016.

[27] S. Han, T. J. Seok, N. Quack, B.-W. Yoo, and M. C. Wu, "Monolithic 50x50 MEMS Silicon Photonic Switches with Microsecond Response Time," in Proc. OFC/NFOEC, San Francisco, CA, pp. M2K.2, Mar. 2014.

[28] L. Xu, W. Zhang, Q. Li, J. Chan, H. L. R. Lira, M. Lipson, and K. Bergman, "40-Gb/s DPSK data transmission through a silicon microring switch," IEEE Photonics Technol. Lett., vol. 24, no. 6, pp. 473–475, Mar. 2012.

[29] X. Zhu, Q. Li, J. Chan, A. Ahsan, H. L. R. Lira, M. Lipson, and K. Bergman, "4x44 Gb/s packet-level switching in a second-order microring switch," IEEE Photonics Technol. Lett., vol. 24, no. 17, pp. 1555–1557, Sep. 2012.

[30] B. G. Lee, B. A. Small, K. Bergman, Q. Xu, M. Lipson, "Transmission of high-data-rate optical signals through a micrometer-scale silicon ring resonator," Opt. Lett. vol. 31, no. 18, pp. 2701-2703, Sep. 2006.

[31] A. Biberman, N. Sherwood-Droz, B. G. Lee, M. Lipson, and K. Bergman, "Thermally active 4×4 non-blocking switch for networks-on-chip," in IEEE 21st Annual Meeting of Lasers and Electro-Optics Society (LEOS), paper 370–371, 2008.

[32] A. Gazman, C. Browning, M. Bahadori, Z. Zhu, P. Samadi, S. Rumley, V. Vujicic, L. P. Barry, K. Bergman, "Software-Defined Control-Plane for Wavelength Selective Unicast and Multicast of Optical Data in a Silicon Photonic Platform," Opt. Express, vol. 25, no. 1, pp. 232-242, Jan. 2017.

[33] L. Chen, N. Sherwood-Droz, and M. Lipson, "Compact bandwidth-tunable microring resonators," Opt. Lett., vol. 32, no. 22, pp. 3361-3362, Nov. 2007.

[34] M. A. Popović, T. Barwicz, M. R. Watts, P. T. Rakich, M. S. Dahlem, F. Gan, C. W. Holzwarth, L. Socci, H. I. Smith, E. P. Ippen and F. X. Kärtner, "Strong-Confinement Microring Resonator Photonic Circuits," in Proc. Lasers and Electro-Optics Society (LEOS), paper TuCC3, 2007.

[35] T. Dai, A. Shen, G. Wang, Y. Wang, Y. Li, X. Jiang, and J. Yang, "Bandwidth and wavelength tunable optical passband filter based on silicon multiple microring resonators," Opt. Lett., vol. 41, no. 20, pp. 4807-4810, Oct. 2016.

[36] V. López, L. Velasco, "Elastic Optical Networks: Architectures, Technologies, and Control," Springer, Jun 2016.

[37] Fujitsu, "CDC ROADM Applications and Cost Comparison" OFC Exhibitor White Papers, 2016.

[38] P. Roorda, "CDC and gridless ROADM architectures and their enabling devices," ECOC 2012 Market Focus, Amsterdam, Netherlands, 2012.

[39] F. Testa, C. J. Oton, C. Kopp, J.-M. Lee, R. Ortuno, R. Enne, S. Tondini, G. Chiaretti, A. Bianchi, P. Pintus, M.-S. Kim, D. Fowler, J. A. Ayucar, M. Hofbauer, M. Mancinelli, M. Fournier, G. B. Preve, N. Zecevic, C. L. Manganelli, C. Castellan, G. Pares, O. Lemonnier, F. Gambini, P. Labeye, M. Romagnoli, L. Pavesi, H. Zimmermann, F. Di Pasquale, and S. Stracca, "Design and Implementation of an Integrated Reconfigurable Silicon Photonics Switch Matrix in IRIS Project," IEEE Sel. Top. Quantum Electron., vol. 22, no. 6, pp. 3600314, Nov/Dec. 2016.

[40] A. Deore, "What comes after 100G? DWDM Super-channels," Infinera,Available:http://www.comsocscv.org/docs/Superchannels_Mar2012.pdf.

[41] H. Furukawa, J. M. Delgado Mendinueta, N. Wada, and H. Harai, "Spatial and Spectral Super-channel Optical Packet Switching System for Multigranular SDM-WDM Optical Networks," IEEE/OSA J. Opt. Commun. Networking, vol. 9, no. 1, pp. A77-A84, Jan. 2017.

[42] X. Tan, M. Yang, L. Zhang, Y. Jiang, and J. Yang, "A Generic Optical Router Design for Photonic Network-on-Chips," J. Lightwave Technol., vol. 30, no. 3, pp. 368-376, Feb. 2012.

[43] R. Stabile, P. DasMahapatra, K. A. Williams, "4×4 InP Switch Matrix with Electro-Optically Actuated Higher-Order Micro-Ring Resonators," IEEE Photonics Technol. Lett. DOI: 10.1109/LPT.2016.2625746, Nov. 2016.

[44] K. Padmaraju, J. Chan, L. Chen, M. Lipson, and K. Bergman, "Thermal stabilization of a microring modulator using feedback control," Opt. Express, vol. 20, no. 27, pp. 27999–28008, Sep. 2012.

[45] M. D. G. Pascual, R. Zhou, F. Smyth, P. M. Anandarajah, and L. P. Barry, "Software reconfigurable highly flexible gain switched optical frequency comb source," Opt. Express, vol. 23, no. 18, pp. 23225–23235, Aug. 2015.

[46] R. Schmogrow, B. Nebendahl, M. Winter, A. Josten, D. Hillerkuss, S. Koenig, J. Meyer, M. Dreschmann, M. Huebner, C. Koos, J. Becker, W. Freude, and J. Leuthold, "Error Vector Magnitude as a Performance Measure for Advanced Modulation Formats," IEEE Photonics Technol. Lett., vol. 24, no. 1, pp. 61-63, Jan. 2012.